\documentstyle[aps,prl,twocolumn,epsf]{revtex}
\begin{document}
\title{Chaos and synchronized chaos in an earthquake model}
\author{Maria de Sousa Vieira\cite{email}}
\address{Department of Biochemistry and Biophysics, University of California, 
San Francisco, CA 94143-0448, USA.}
\maketitle
\begin{abstract}
We show that chaos is present in the symmetric two-block  
Burridge-Knopoff model for earthquakes. This is in 
contrast with previous numerical studies, but in 
agreement with experimental results. In this system,  
we have found a rich dynamical behavior with an unusual route 
to chaos. 
In the three-block system, we see 
the appearance of synchronized chaos, 
showing that this concept can have potential 
applications in the field of seismology. 
\end{abstract} 
\narrowtext   
In recent years, the phenomenon of chaotic synchronization 
has been a subject of intensive studies.   
By definition, chaotic systems present strong sensitivity 
to the initial conditions, and in principle it seems impossible to  
synchronize them. However, 
Fujisaka {\sl et al.}\cite{fujisaka} 
and Pecora {\sl et al.}\cite{pecora} showed 
that systems  
with chaotic behavior can be synchronized, if 
appropriate connections among them are made.  
This phenomenon has been called ``chaotic synchronization", 
and has been investigated mainly in applications 
for secure communications\cite{dpll}. 

Another area of active research nowadays is related to 
systems that present avalanche-like dynamics.  
This was triggered by a paper 
by Bak, Tang and Wiesenfeld\cite{soc}. They showed that certain 
dissipative systems, with many degrees of freedom, naturally 
evolve to a critical state characterized by power-law distributions 
in space and time. They denoted this phenomenon Self-Organized 
Criticality (SOC).

One of the systems that have been studied in connection with
SOC is the Burridge-Knopoff (B-K) model for earthquakes\cite{bk}.
This model consists of blocks connected by springs. The whole
systems is pulled with constant velocity on a surface with
friction. It has been shown experimentally\cite{bk} and numerically\cite{cl}
that this model presents a region of power-law distribution
similar to what is observed in real earthquakes, namely,
the Gutenberg-Richter law\cite{gr}. Since the power-law
does not span the entire system, one could say that
this system does not present what has been defined as SOC.
However, a variation of it, called the ``train model", does present
SOC\cite{train,chaos}. 

After the work by Carlson and Langer\cite{cl}, several studies 
on the B-K model were performed. 
With respect to the chaotic properties of the model, we 
are aware of the numerical studies by Nussbaum and Ruina\cite{ruina}, 
Huang and Turcotte\cite{huang}, Nakanishi\cite{naka},  
Crisanti {\sl et al.}\cite{crisanti} and Lacorata {\sl et al.}\cite{lacorata}.  
In \cite{naka} and 
\cite{crisanti} the systems considered had more than 
two blocks and they were evolved by cellular automaton rules. 
Nussbaum {\sl et al.} studied a symmetric two-block B-K model, and 
verified that, if a friction force of the Coulomb type (that is 
the dynamic friction coefficient being constant) the system 
presents only period behavior.   
Huang and Turcotte,  and Lacorata and Paladin found chaotic behavior  
in the two-block B-K model only with the presence of
an asymmetry in the system, even considering a  
velocity weakening friction force\cite{huang,lacorata}.  
In particular, they considered the friction 
force in one block being different from the
friction force in the other block.  
On the other hand, 
by modeling the two-block B-K model by electronic 
circuits, Field, Venturi and Nori\cite{nori} showed experimentally that 
a completely {\sl symmetric} system does present chaotic 
behavior in a wide range of the parameter space.  
Therefore, their 
results are in contradiction to what was reported in\cite{huang,lacorata}.  
One of the purposes of this letter is to resolve this contradiction. We show 
that the two-block B-K system in a symmetric configuration is chaotic. 
As in the experimental study, chaos is 
seen in wide range of parameter values. We stress that a one-block 
system in the BK model (with a linear drive)  cannot present chaos, since 
its dimensionality is smaller than the minimum dimension necessary 
for a system to present chaotic behavior, which is three.  
(It is obvious that here 
we are considering the absence of elements with delay in the system, 
since in this way one could increase its dimensionality up 
to infinity). 

In this letter we also show that the  phenomenon of 
synchronized chaos appears in the three-block system of the  
Burridge-Knopoff model for earthquakes.  
Most importantly, it comes naturally from the 
geometry of the system, without any need for special connections, as 
it is generally the case in the studies of synchronized chaos.  
From our results, we speculate that 
synchronized chaos may have applications in the field of seismology.   
That is, earthquakes faults, which are generally coupled 
through the elastic media in the Earth crust could in principle synchronize, 
even when they have an irregular (chaotic) dynamics.
As a consequence of synchronization, the dimensionality of the system 
decreases what simplifies the analysis of the system, 
as discussed below. 
We quantify the degree of synchronization by studying 
the Liapunov exponent associated with the 
synchronization manifold, that is, the 
transverse (or conditional) Liapunov exponent\cite{fujisaka,pecora1}
and compare it with the Liapunov 
exponents of the three-block system. 

We start by reviewing the B-K model. It 
consists of a chain of 
blocks of mass $m$, connected by coil springs of strength $k_c$
to its nearest neighbors. They are situated on a rough surface. 
Between the blocks and the surface there is a 
frictional force  $F$. 
Here we consider that $F$ is a function of the block velocity. 
Each block is also attached by a leaf spring of strength $k_p$ to
a surface that moves with constant velocity $v$. 
A figure of the system can be found in \cite{cl}.
Following 
Carlson and Langer\cite{cl},   
we denote by $X_j$ the position of block 
$j$ with respect to its equilibrium position,  
write the friction force as  
$F(\dot X_j/v_c)= F_{\circ}\Phi(\dot X_j/v_c) $  
where $\Phi (0)=1$ and $v_c$ is a characteristic velocity,   
and introduce the variables 
$\tau \equiv \omega _p t$, $\omega _p^2 \equiv k_p/m$, 
$U_j \equiv k_pX_j/F_{\circ}$. 
Then, the equation of motion for the two-block system can be 
written in the following dimensionless form
\begin{eqnarray}
\ddot U_1 &=& k(U_{2} - U_1) -U_1 + \nu \tau -\Phi(\dot U_1/\nu^c_1), \cr 
\ddot U_2 &=& k(U_{1} -U_2) -U_2 + \nu \tau -\Phi(\dot U_2/\nu^c_2), 
\label{eq1}
\end{eqnarray}
with $\nu \equiv v/V_{\circ}$, 
$\nu^c \equiv v_c/V_{\circ}$, $V_{\circ} \equiv F_{\circ}/\sqrt{k_pm}$ 
and $k \equiv k_c/k_p$. 
Dots denote differentiation with respect to $\tau $. 
Eq.~\ref{eq1} is valid only when block $j$ is moving. If it
is not, its equation is simply $\dot U_j=\nu$.
We use 
the velocity weakening friction force introduced in \cite{cl}, 
given by    
\begin{equation}
\Phi (\dot U/\nu^c)= {{1}\over{1+\dot U/\nu^c}},   
 \label{eq22}
\end{equation}
which is a simple nonlinear function. In the simulations 
displayed here, we did not allow backward motion, that 
is the static friction force can take any value to
prevent it. However, several tests showed that 
if backward motion is allowed, the results remain 
essentially the same. 

In contradiction with the reports in\cite{huang,lacorata}, 
but in agreement with experimental results\cite{nori}, we find that  
a velocity weakening friction force leads to a rich dynamical 
behavior in the two-block B-K system, even when it is identical for 
the two blocks. 
We find the presence of 
periodic, quasiperiodic and chaotic behavior. In  Fig.~\ref{f1}(a) 
we show an 
example of chaotic motion, by   
plotting $\dot U_1$ versus $U_1 - U^e_1$,  
with $1/\nu^c \equiv 1/\nu^c_1 = 1/\nu^c_2 = 1$. We denote  
$U^e_j \equiv \nu \tau - \nu^c/(\nu^c+\nu)$ 
the unstable equilibrium point around which the 
orbits of block $j$ circle in phase space, which is found by 
taking $\ddot U_j=0$ and $\dot U_j=\nu $ in    
Eq.~\ref{eq1} (the stability of such a solution for any 
number of blocks was analyzed in\cite{cl}). 
Unless explicitly 
stated otherwise, we take in the numerical studies shown here $k=1$ and 
$\nu = 0.1$. However, similar behavior was found for other 
parameter values, as well. We do not display  
$\dot U_2$ versus $U_2 - U_2^e$, since we have found that    
for this system   
the attractors of the two blocks are the same. In other words, 
the plot of $\dot U_2$ versus $U_2 - U_2^e$ is identical to the 
one of $\dot U_1$ versus $U_1 - U_1^e$, showing that  
the two blocks will visit the same region of the phase-space, 
but not necessarily at the same time.
The initial conditions for the simulations shown here 
are the blocks initially at rest and with small random displacements 
from their equilibrium position.  
In all the simulations, a transient time  
was discarded.  
    
The bifurcation diagram for the two-block system is shown 
in Fig.~\ref{f1}(b). There, we display $\dot U_1$ versus $1/\nu^c$, in the 
Poincare section satisfying $U_1-U^e_1=0$. 
In order to quantitatively characterize the dynamics, we have 
calculated the two largest Liapunov exponents of the system. 
If the largest Liapunov exponent (LLE) of the system is greater  
than zero, then, by definition the system is chaotic. 
Quasiperiodic motion occurs in this system when 
the  LLE and the second largest 
Liapunov exponent (SLLE) are zero, and 
the motion is periodic when the LLE is zero and the SLLE 
is negative.  We used the method introduced in \cite{liap} to 
calculate the LLE and SLLE, and our results 
are displayed in Fig.~\ref{f1}(c). 
The solid line refers to the LLE and the dashed line is the SLLE.   
We investigated in detail the first entrance into chaos, which 
occurs at $1/\nu ^c \approx 0.112$  
for these parameter values, and noticed that it is  
unusual.  
We found that the route to chaos is from period one, to 
period two and then directly into chaos.  
This unusual route is probably due to the fact that here 
we have a system governed by 
non differentiable flows. Most of what is known 
in dynamical system theory deals with systems that are 
infinitely differentiable, which is not the case here.  
More details about return maps and other quantities for 
this system will be published elsewhere\cite{unp}. 

One may ask, why Huang and Turcotte\cite{huang} and Lacorata and 
Paladin\cite{lacorata} 
did not find chaotic behavior 
in the symmetric B-K model? The reason is that  
they made an inconsistent assumption,  
that is, that the driving block does not move during 
the slipping events. This assumption is equivalent to take 
the pulling velocity going to zero, since they drop out 
the term $\nu \tau$ in 
Eq.~\ref{eq1}. 
However, from the equations  
of motion of the system, it turns out that the blocks will not move, if they 
are initially at rest,  
if one considers this, since the displacement of the blocks is proportional 
to the pulling velocity. This is shown in detail in\cite{cl}.
On the other hand, the blocks will not stop if they 
are initially in motion. That is, stick-slip dynamics does not occur 
when the system is evolved by the equation used by those authors.   
One way to avoid this 
is to take a discontinuous friction force as in\cite{cl2}, 
and this was not considered in the study of 
the symmetric system by Huang and Turcotte, and Lacorata and Paladin.
In other words, their equations of motion does not describe the 
dynamics of the Burridge-Knopoff model.

Now, we  concentrate on the phenomenon of chaotic synchronization 
in the B-K model. We have not found this phenomenon  
in the two-block system with the friction force we consider 
here. If the two blocks were synchronized, they would behave as 
a single block, and this would be equivalent to have a single 
block system with chaotic behavior, which we know it is impossible, as 
discussed above. We see however, that a three block system does 
present chaotic synchronization. 

The equation of motion for the three-block system in dimensionless form is 
\begin{eqnarray}
\ddot U_1 &=& k(U_{2} - U_1) -U_1 + \nu \tau -\Phi(\dot U_1/\nu^c_1),  \cr 
\ddot U_2 &=& k(U_{1} -2U_2+U_{3}) -U_2 + \nu \tau -\Phi(\dot U_2/\nu^c_2), \cr
\ddot U_3 &=& k(U_{2} -U_3) -U_3 + \nu \tau -\Phi(\dot U_3/\nu^c_3). 
\label{eq5}
\end{eqnarray}
We see in the equations that govern the motion of  blocks 1 and 3 have the  
same functional form. They are also linked to a common 
subsystem, that is, block 2. This configuration is not 
of the ``master-slave" type, since there is feedback 
between blocks 1 and 2 and between blocks 2 and 3. However, the 
equations have  
the necessary ingredients for chaotic synchronization between blocks 1 and 3 to 
occur, which 
is the same functional form. 
 
We have found chaotic synchronization when the 
parameters for all the blocks are the same only in a 
very small region of the parameter space. It is 
not difficult to understand why. With the leaf 
springs having the same value,  blocks 1 and 
3 are more loose than block 2 (since the later 
is attached to two coil springs instead of one). Therefore, 
in general,  
blocks 1 and 3 attain larger velocities than 
block 2,  and with the friction force that we
use, they are more unstable. Chaotic synchronization 
happens when 
the subsystems to be synchronized are  
be more stable than the subsystem to which they are 
connected\cite{pecora}.   
Also, it is a property of chaotic synchronization that  
blocks 1 and 3 need to have the same parameter 
values, if perfect synchronization in the 
absence of control is the goal.   

However, we find that when there is  an asymmetry in the system, chaotic 
synchronization occurs in a large range of the parameter space. 
We can either introduce asymmetries in the springs, masses, 
or friction forces. We choose the later to demonstrate 
our results, but similar results were found in the 
other two cases. Therefore, we make the friction 
force in block 2 smaller than the friction force 
in blocks 1 and 3 (which means that they are 
more rough than block 2).  

In Fig.~\ref{f2}(a) we show an example of a chaotic orbit 
for block 1, with the parameter values 
$1/\nu^c \equiv 1/\nu^c_1 = 1/\nu^c_3 = 4/\nu^c_2=0.165$. 
We will see that this attractor is from a regime in which  
blocks 1 and 3 are synchronized.    
In Fig.~\ref{f2}(b) the LLE (solid line) and 
the SLLE (dashed line) of the three-block 
system is shown.
In  Fig.~\ref{f2}(c) we show the Liapunov exponent of 
the synchronization manifold (solid line), that is, 
what is called the transverse (or conditional) 
Liapunov exponent\cite{fujisaka,pecora1}. 
We have calculated the transverse 
Liapunov exponent by adapting the method introduced 
by Benettin\cite{liap}. That is, after the transient 
dies out, we evolve the orbit 
of block 3 by 
making it slightly different from the orbit of block 1. 
Then, we verify how the difference between 
the orbits of the two blocks      
evolves after a short time step. 
The perturbation is renormalized 
in the direction of the maximum growth, 
and the process is repeated many times. 
The transverse Liapunov exponent is given 
by the average  logarithm (in this paper we use base 2) 
of the 
growth (or shrinkage) of the perturbation along the orbit.  
Fig.~\ref{f2}(c) also shows (dashed line)  
the Euclidean  distance $D$ in phase-space 
between blocks 1 and 3, that is, 
$D \equiv \sqrt{(U_1 - U_3)^2 + (\dot U_1 - \dot U_3)^2}$, 
as a function of 
$1/\nu^c$. 
This distance is an average over a time 
$\Delta \tau=2000$. 
We see that the transverse Liapunov exponent correctly 
determines the region in which blocks 1 and 3 are synchronized.
In this situation, the transverse Liapunov exponent is negative and 
the distance between the two blocks is zero. 
Comparing Fig.~\ref{f2}(b) and Fig.~\ref{f2}(c) we can 
identify the regions of chaotic synchronization, where 
we have one or more Liapunov exponents 
of the three-block system greater than zero and the transverse Liapunov 
exponent is less than zero. 
This is, for example, the case of the attractor shown 
in Fig.~\ref{f2}(a). There, only one Liapunov exponent of the 
system is larger than zero. 
The comparison between Fig.~\ref{f2}(b) and Fig.~\ref{f2}(c) 
also shows that the transition from chaos to hyperchaos\cite{hyper}
does not determine here  
the transition from chaotic synchronization 
to non-synchronization, as it was the case of the system 
studied in\cite{mll} (the hyperchaos regime is defined 
as the one in which there is more than one positive Liapunov 
exponent). 
For example, at $1/\nu^c =0.38$ there is only one positive 
Liapunov exponent, and no synchronization is seen between 
blocks 1 and 3. There is also the case in which 
the LLE and the SLLE are non-positive, and 
blocks 1 and 3 do not synchronize, as in $1/\nu^c =0.4$. 

Now we discuss what is the possible relevance of our finds to 
the analysis of real earthquakes. 
It is known that earthquakes are not strictly periodic 
phenomena. 
If their irregular 
behavior is caused by  
a chaotic, deterministic process, 
as this simple spring-block model suggests, then 
in principle, prediction about them   
could be made in a short time scale. 
In this way, 
the damage  they 
cause could be minimized. 
The problem with modeling chaotic systems is that,  
if the dimension of their attractor is very large, 
very little can be done with respect to prediction\cite{farmer}. 
Such systems are treated basically as stochastic systems.
However, if synchronization happens among 
elements of a large system, then the dimensionality 
of the attractor decreases. For example, in the 
three-block system studied here, the dimension 
of the attractor can decrease by up to two. If the 
dimension of the attractor decreases, the analysis 
of the system becomes easier. This fact emphasizes the 
need for a deeper analysis of the relationship 
between synchronized chaos  
and prediction. 
We hope that this work will motivate more studies on chaotic  
synchronization for applications to seismology.

\begin{figure}
\caption[f1]{
Orbit in phase-space for block 1 when  
$1/\nu^c_1=1/\nu^c_2=1$.  
(b) Bifurcation diagram of $\dot U_1$ in the Poincare section in which 
$U_1-U^{e}_1=0$. (c) The largest Liapunov exponent $\lambda _1$ (solid line)
and the second largest Liapunov exponent $\lambda _2$ (dashed line) of the system. Here we 
have $\nu =0.1$ and $k=1$ in a two-block system. The Liapunov exponents were calculated with 
perturbations of $10^{-4}$, time step of $0.05$, and with  $400000$ iterations.
The dotted line at $\lambda =0 $ is 
just a guide to the eye.} 
\label{f1}
\end{figure}

\begin{figure}
\caption[f2]{
Orbits in phase-space for $\dot U_1$ versus $U_1 - \nu \tau$ when   
$1/\nu^c \equiv 1/\nu^c_1=1/\nu^c_3=4/\nu^c_2=0.165$.       
(b) The largest Liapunov exponent $\lambda _1$ (solid)
and the second largest Liapunov exponent $\lambda _2$ 
(dashed) as a function of 
$1/\nu^c$.  (c) The transverse Liapunov exponent $\lambda _t$
(solid line) and the  
Euclidean distance $D$ in phase-space 
between blocks 1 and 
3 (dashed line) a function of $1/\nu^c $. Here we have  
$\nu =0.1$, $k=1$ in a three-block system. 
The Liapunov exponents were calculated with 
perturbations of $10^{-4}$, time step of $0.05$, and with $400000$ iterations.
The dotted line at $D=0$ is just a 
guide to the eye.} 
\label{f2}
\end{figure}

\newpage
\input psfig
\noindent\vbox{
\vskip 1.50truecm
\centerline{\psfig{file=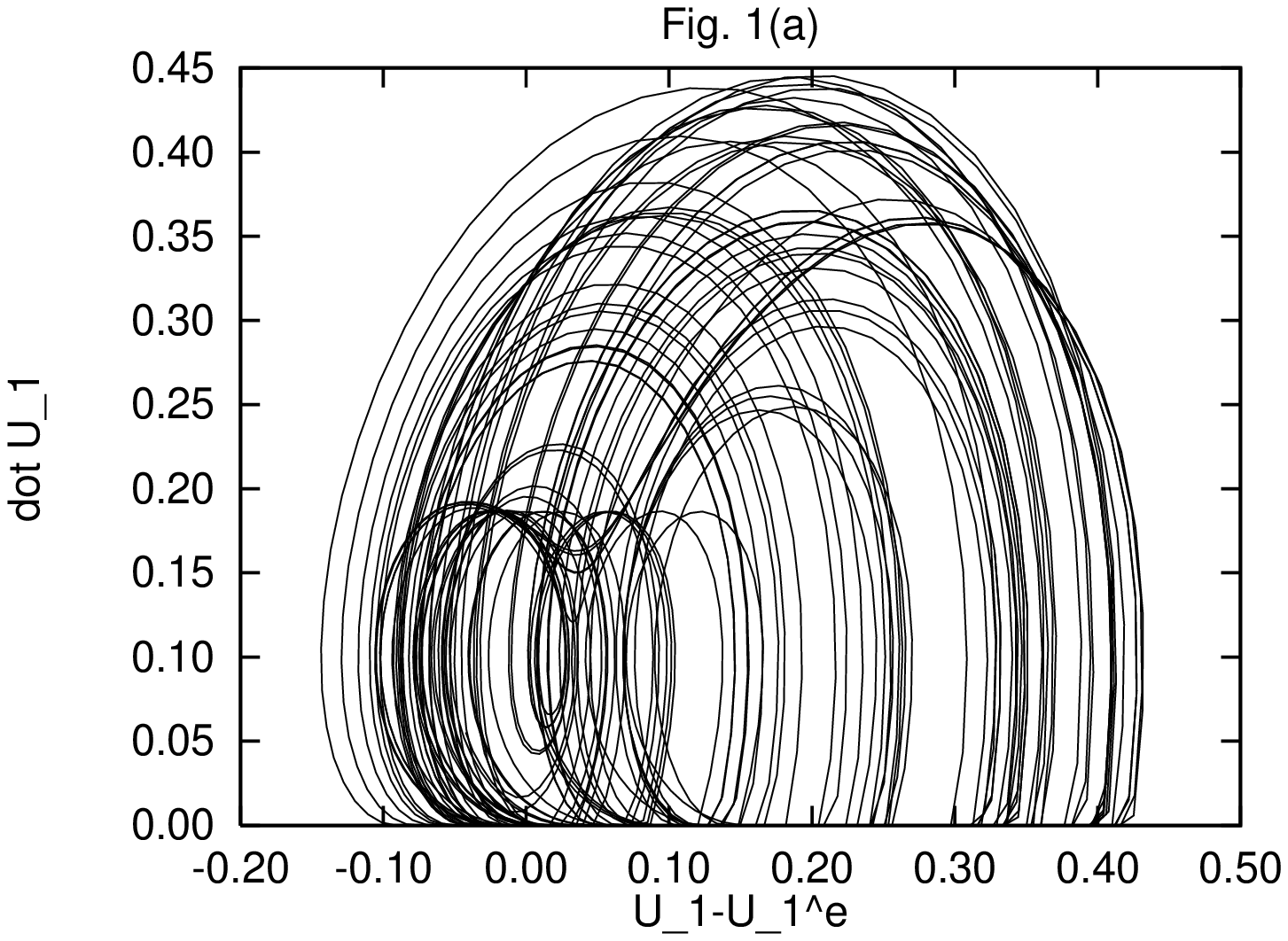,height=5cm}}
\vskip 0.0truecm
}
\noindent\vbox{
\vskip 1.5truecm
\centerline{\psfig{file=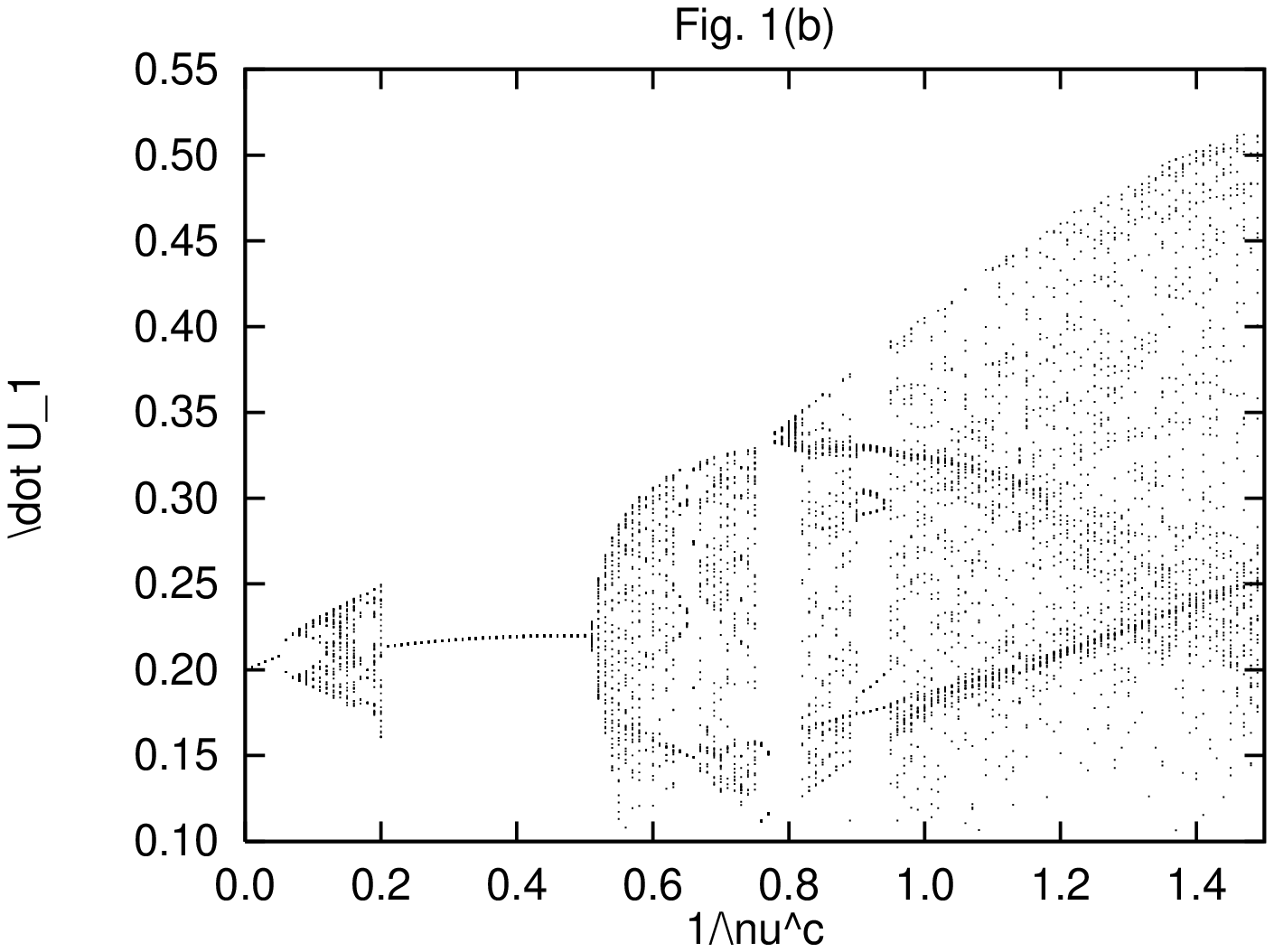,height=5cm}}
\vskip 0.0truecm
}
\noindent\vbox{
\vskip 1.5truecm
\centerline{\psfig{file=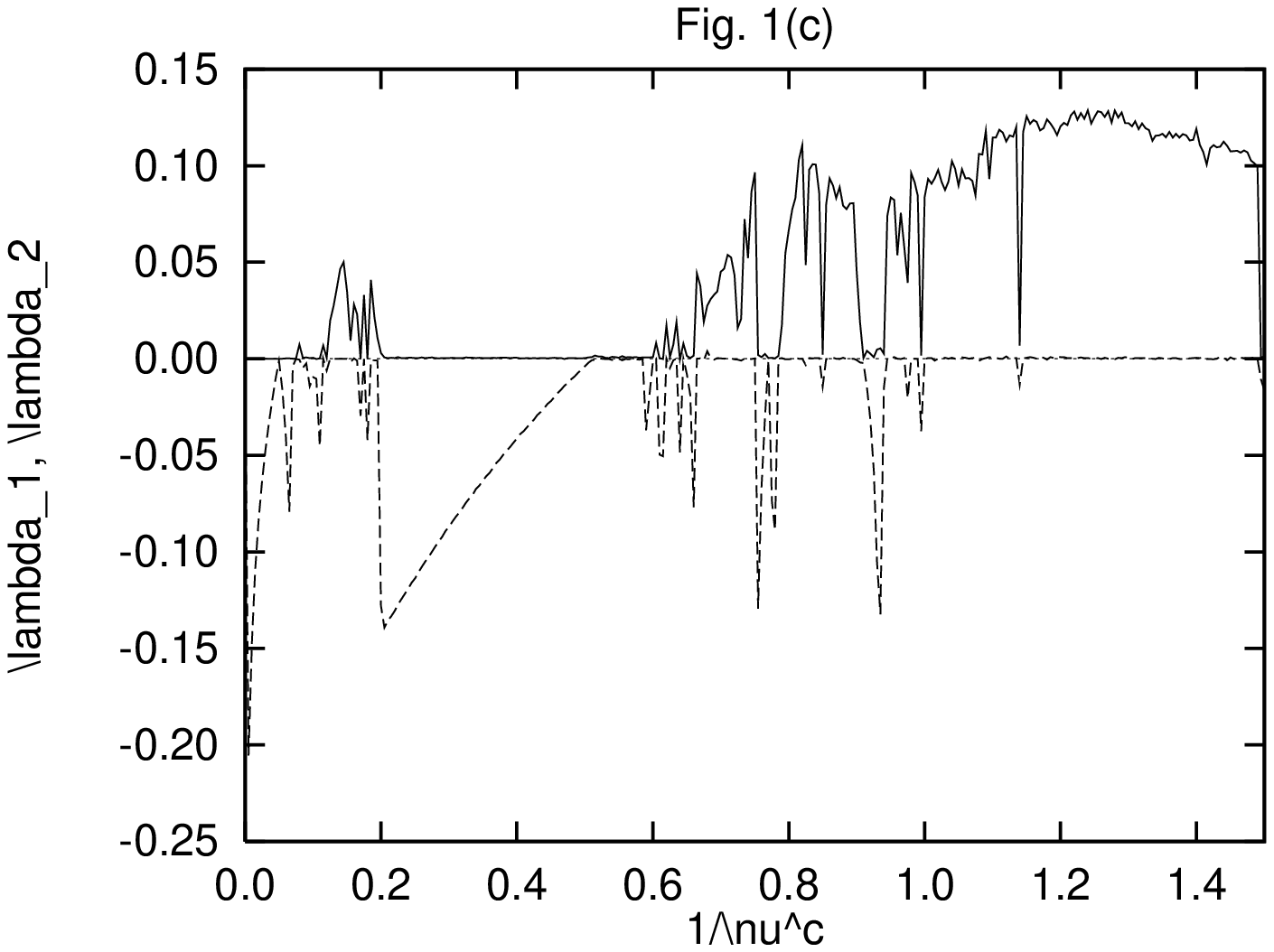,height=5cm}}
}

\newpage
\noindent\vbox{
\vskip 1.50truecm
\centerline{\psfig{file=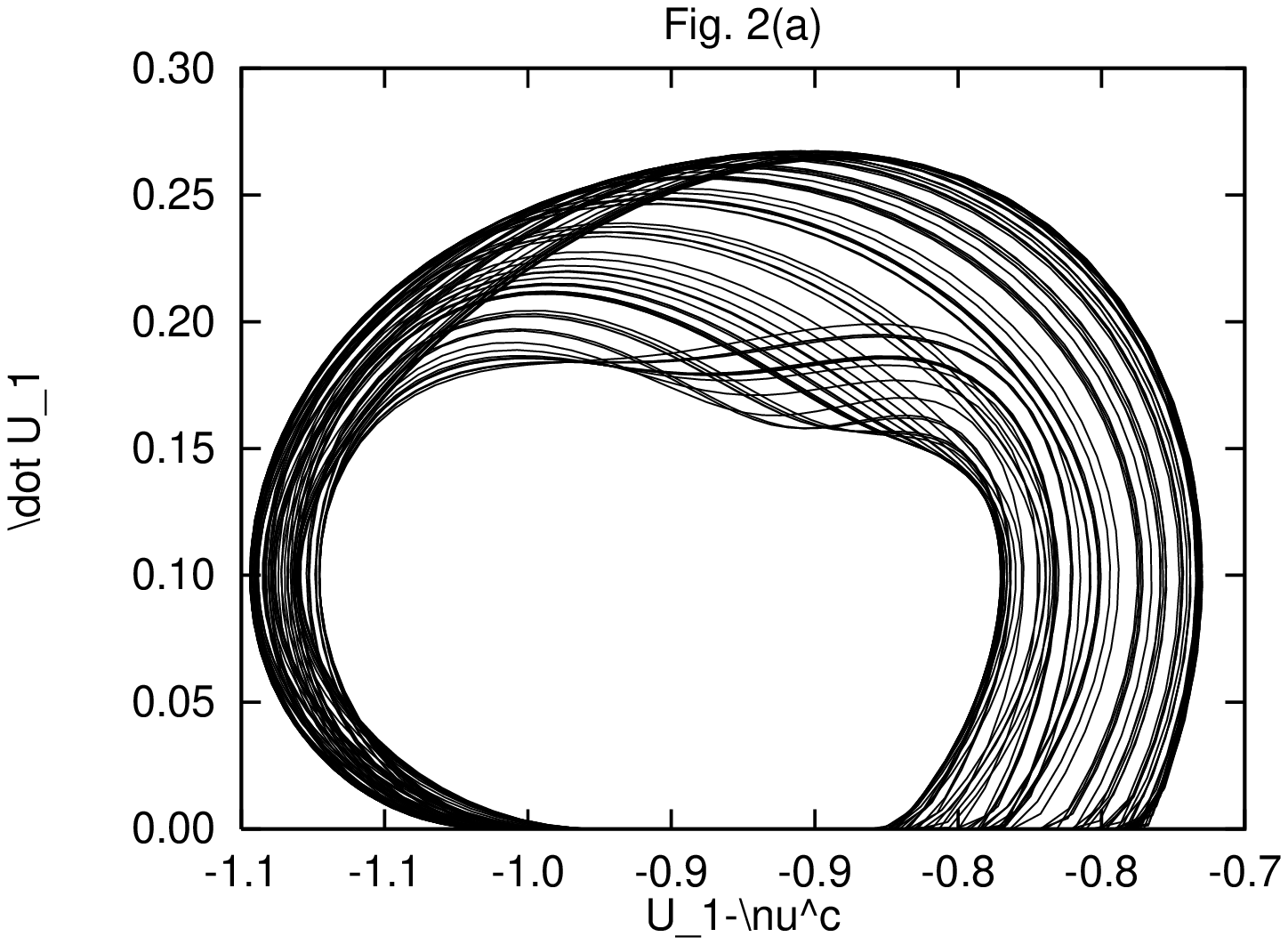,height=5cm}}
\vskip 0.0truecm
}
\noindent\vbox{
\vskip 1.5truecm
\centerline{\psfig{file=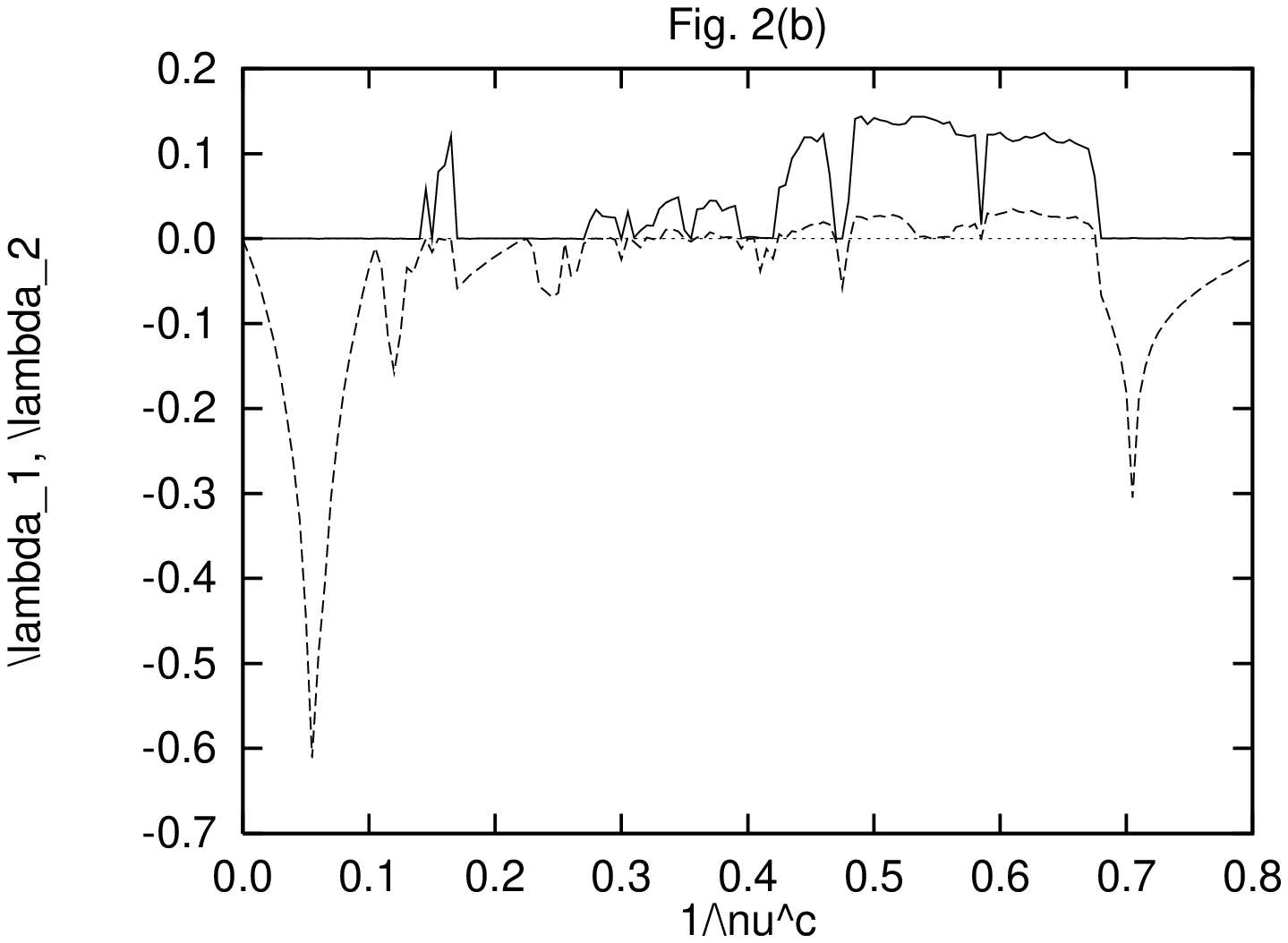,height=5cm}}
\vskip 0.0truecm
}
\noindent\vbox{
\vskip 1.5truecm
\centerline{\psfig{file=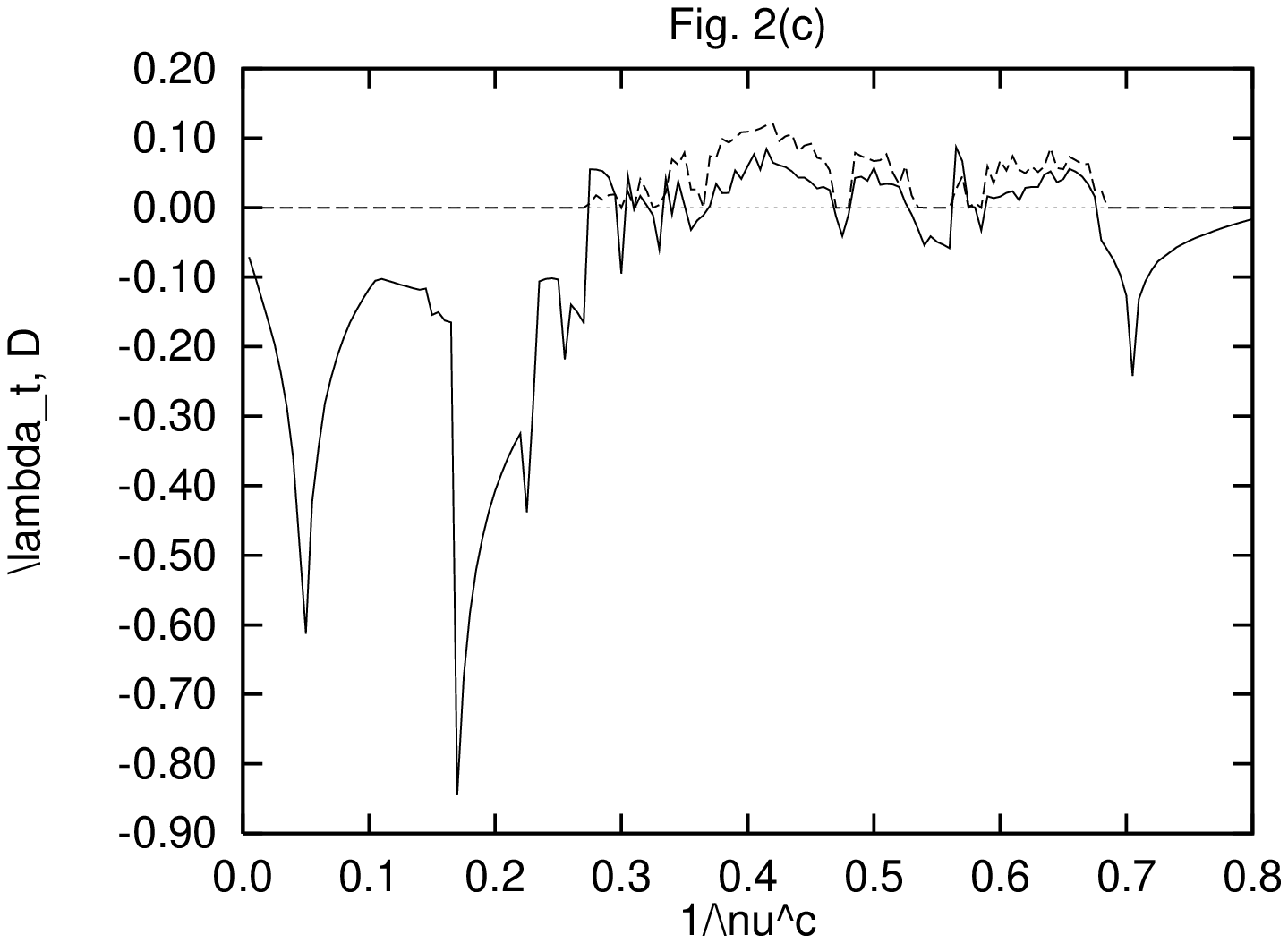,height=5cm}}
}

\end{document}